\RequirePackage{fix-cm}
\documentclass[twocolumn,epjc3-upd,fleqn]{svjour3}
\smartqed  
\RequirePackage{graphicx}
\RequirePackage{subfigure}
\usepackage{rotating}
\usepackage{amsmath}
\usepackage{mathtools}
\RequirePackage[colorlinks,citecolor=blue,urlcolor=blue,linkcolor=blue]{hyperref}
\usepackage{doi}
\usepackage[]{units}
\usepackage{xspace}
\usepackage[switch]{lineno}
\usepackage{cite}
\usepackage{comment}
\usepackage{url}

\newcommand{\DBD}{0$\nu\beta\beta$}
\newcommand{\LMO}{Li$_{2}${}$^{100}$MoO$_4$}

\newcommand{\CupidZ}{CUPID-0}

\begin{document}

\title{Characterization of cubic Li$_{2}${}$^{100}$MoO$_4$ crystals for the CUPID experiment.}

\author{A.~Armatol\thanksref{CEA_IRFU_France}
\and
E.~Armengaud\thanksref{CEA_IRFU_France}
\and
W.~Armstrong\thanksref{ANL_US}
\and
C.~Augier\thanksref{IP2I_France}
\and
F.~T.~Avignone~III\thanksref{UofSC_US}
\and
O.~Azzolini\thanksref{LNL_Italy}
\and
A.~Barabash\thanksref{ITEP_Russia}
\and
G.~Bari\thanksref{SdB_Italy}
\and
A.~Barresi\thanksref{MIB_Italy,UniMIB_Italy}
\and
D.~Baudin\thanksref{CEA_IRFU_France}
\and
F.~Bellini\thanksref{SdR_Italy,SURome_Italy}
\and
G.~Benato\thanksref{LNGS_Italy}
\and
M.~Beretta\thanksref{UCB_US}
\and
L.~Berg\'e\thanksref{IJCLab_France}
\and
M.~Biassoni\thanksref{MIB_Italy}
\and
J.~Billard\thanksref{IP2I_France}
\and
V.~Boldrini\thanksref{CNR-IMM_Italy,SdB_Italy}
\and
A.~Branca\thanksref{MIB_Italy,UniMIB_Italy}
\and
C.~Brofferio\thanksref{MIB_Italy,UniMIB_Italy}
\and
C.~Bucci\thanksref{LNGS_Italy}
\and
J.~Camilleri\thanksref{VT_US}
\and
S.~Capelli\thanksref{MIB_Italy,UniMIB_Italy}
\and
L.~Cappelli\thanksref{LNGS_Italy}
\and
L.~Cardani\thanksref{SdR_Italy}
\and
P.~Carniti\thanksref{MIB_Italy,UniMIB_Italy}
\and
N.~Casali\thanksref{SdR_Italy}
\and
A.~Cazes\thanksref{IP2I_France}
\and
E.~Celi\thanksref{LNGS_Italy,GSSI}
\and
C.~Chang\thanksref{ANL_US}
\and
M.~Chapellier\thanksref{IJCLab_France}
\and
A.~Charrier\thanksref{CEA_IRAMIS_France}
\and
D.~Chiesa\thanksref{MIB_Italy,UniMIB_Italy}
\and
M.~Clemenza\thanksref{MIB_Italy,UniMIB_Italy}
\and
I.~Colantoni\thanksref{SdR_Italy,CNR-NANOTEC}
\and
F.~Collamati\thanksref{SdR_Italy}
\and
S.~Copello\thanksref{SdG_Italy,UnivGenova}
\and
F.~Cova\thanksref{MIB_Italy,UniMIB_Italy}
\and
O.~Cremonesi\thanksref{MIB_Italy}
\and
R.~J.~Creswick\thanksref{UofSC_US}
\and
A.~Cruciani\thanksref{SdR_Italy}
% ...
}

\institute{IRFU, CEA, Universit\'e Paris-Saclay, Saclay, France\label{CEA_IRFU_France}
\and
Argonne National Laboratory, Argonne, IL, USA\label{ANL_US}
\and
Institut de Physique des 2 Infinis, Lyon, France\label{IP2I_France}
\and
University of South Carolina, Columbia, SC, USA\label{UofSC_US}
\and
INFN Laboratori Nazionali di Legnaro, Legnaro, Italy\label{LNL_Italy}
\and
National Research Centre Kurchatov Institute, Institute for Theoretical and Experimental Physics, Moscow, Russia\label{ITEP_Russia}
\and
INFN Sezione di Bologna, Bologna, Italy\label{SdB_Italy}
\and
INFN Sezione di Milano - Bicocca, Milano, Italy\label{MIB_Italy}
\and
University of Milano - Bicocca, Milano, Italy\label{UniMIB_Italy}
\and
INFN Sezione di Roma, Rome, Italy\label{SdR_Italy}
\and
Sapienza University of Rome, Rome, Italy\label{SURome_Italy}
\and
INFN Laboratori Nazionali del Gran Sasso, Assergi (AQ), Italy\label{LNGS_Italy}
\and
University of California, Berkeley, CA, USA\label{UCB_US}
\and
Universit\'e Paris-Saclay, CNRS/IN2P3, IJCLab, Orsay, France\label{IJCLab_France}
\and
CNR-Institute for Microelectronics and Microsystems, Bologna, Italy\label{CNR-IMM_Italy}
\and
Virginia Polytechnic Institute and State University, Blacksburg, VA, USA\label{VT_US}
\and
Gran Sasso Science Institute, L'Aquila, Italy\label{GSSI}
\and
IRAMIS, CEA, Universit\'e Paris-Saclay, Saclay, France\label{CEA_IRAMIS_France}
\and
CNR-Institute of Nanotechnology, Rome, Italy\label{CNR-NANOTEC}
\and
INFN Sezione di Genova, Genova, Italy\label{SdG_Italy}
\and
University of Genova, Genova, Italy\label{UnivGenova}
% ...
\and
Lawrence Berkeley National Laboratory, Berkeley, CA, USA\label{LBNL_US}
\and
Northwestern University, Evanston, IL, USA\label{NWU_US}
\and
Massachusetts Institute of Technology, Cambridge, MA, USA\label{MIT_US}
\and
INFN Laboratori Nazionali di Frascati, Frascati, Italy\label{LNF_Italy}
% ...
\and
California Polytechnic State University, San Luis Obispo, CA, USA\label{CalPoly_US}
% ...
\and
Yale University, New Haven, CT, USA\label{Yale_US}
\and
University of California, Los Angeles, CA, USA\label{UCLA_US}
\and
Drexel University, Philadelphia, PA, USA\label{Drexel_US}
% ...
\and
Nikolaev Institute of Inorganic Chemistry, Novosibirsk, Russia\label{NIIC_Russia}
\and
Johns Hopkins University, Baltimore, MD, USA\label{JHU_US}
\and
INFN Sezione di Padova, Padova, Italy\label{PD_Italy}
\and
University Grenoble Alpes, CNRS, Grenoble INP, SIMAP, Grenoble, France\label{SIMaP_Grenoble_France}
\and
University of Bologna, Bologna, Italy\label{UnivBologna_Italy}
}

\thankstext{e1}{Also at: Physik-Department, Technische Universit{\"a}t M{\"u}nchen, Garching, Germany}

\date{Received: date / Accepted: date}

\maketitle

\begin{abstract}
The CUPID Collaboration is designing a tonne-scale, background-free detector to search for double beta decay with sufficient sensitivity to fully explore the parameter space corresponding to the inverted neutrino mass hierarchy scenario. 
One of the CUPID demonstrators, CUPID-Mo, has proved the potential of enriched \LMO\ crystals as suitable detectors for neutrinoless double beta decay search. In this work, we characterised cubic crystals that, compared to the cylindrical crystals used by CUPID-Mo, are more appealing for the construction of tightly packed arrays. 
We measured an average energy resolution of (6.7$\pm$0.6)\,keV FWHM in the region of interest, approaching the CUPID target of 5\,keV FWHM. We assessed the identification of $\alpha$ particles with and without a reflecting foil that enhances the scintillation light collection efficiency, proving that the baseline design of CUPID already ensures a complete suppression of this $\alpha$-induced background contribution.
We also used the collected data to validate a Monte Carlo simulation modelling the light collection efficiency, which will enable further optimisations of the detector.

\end{abstract}

\section{Introduction}
\label{intro}
Double beta decay occurs when a nucleus spontaneously changes its atomic number by two units~\cite{GoeppertMayer}. Despite the low probability for this process to happen, double beta decay has been already observed for 12 nuclei, with typical half-lives in the range of 10$^{18}$--10$^{24}$\,yr~\cite{BARABASH2020}. In recent times, some experimental techniques have reached such a high precision on the measurement of this process that today it is possible to infer important nuclear properties or even search for physics beyond the Standard Model of Particles and Fields (SM) by studying spectral distortions~\cite{PhysRevD.100.092002,PhysRevLett.123.262501,PhysRevD.93.072001,nemo3-dbd,PhysRevLett.122.192501,PhysRevD.98.092007,Armengaud2020_2nu}.

Many theoretical frameworks predict that double beta decay can also occur without the emission of neutrinos~\cite{Furry_1939,Deppisch:2012nb}. Such a process, forbidden by the SM, will result in the creation of only two electrons, thus violating the conservation of the total lepton number~\cite{Vissani:2016}. Furthermore, neutrinoless double beta decay (\DBD) will occur if neutrinos and antineutrinos are the same particles, in contrast to all other fermions~\cite{PhysRevD.25.2951}. Thus, the observation of this process would allow the inference of fundamental properties of neutrinos and set important milestones for leptogenesis theories and particle physics.

Despite the tremendous progress in the past few decades, \DBD\ keeps eluding detection. The most competitive experiments are now setting lower limits on its half-life in the range of 10$^{24}$--10$^{26}$\,yr~\cite{Agostini1445,PhysRevLett.117.082503,PhysRevC.100.025501,PhysRevLett.123.161802,PhysRevD.92.072011,PhysRevLett.123.032501,PhysRevLett.124.122501,Armengaud:2020:cupidmo,gerdacollaboration2020final}. 
The proposed next-generation experiment CUPID (CUORE Upgrade with Particle IDentification~\cite{CUPID_preCDR_2019,CUPID_2015,CUPID_RD_2015}) aims to push the half-life sensitivity beyond 10$^{27}$\,yr by operating a tonne-scale detector in background-free conditions.

CUPID will rely on the well established technology of scintillating cryogenic calorimeters (often referred as scintillating bolometers). 
Cryogenic calorimeters have been developed for almost 40 years~\cite{Fiorini:1983yj,pirroreview}, starting from samples of few grams and proving the feasibility of tonne-scale detectors through the CUORE experiment~\cite{Q_AHEP}. The 
significant results obtained by CUORE, which recently reached 1 ton$\cdot$yr exposure,
marks for its successor CUPID an important milestone to build on.

The success of CUPID hinges on an important technological innovation: implementation of a background-free technology. According to the CUORE background model~\cite{cuore-bkg-2017}, the dominant background source for cryogenic calorimeters are $\alpha$ particles produced by the radioactive decays of the residual contamination of the materials constituting the detector. The CUORE Collaboration has already pushed the radiopurity limit with strict material selection and cleaning techniques. A further background suppression can be obtained mainly through particle identification. CUPID will exploit the simultaneous read-out of the calorimetric signal and scintillation light, taking advantage of the different light yield of electrons (potential signal) and $\alpha$ particles, to actively reject the $\alpha$ background~\cite{Bobin:1997qm,Pirro:2005ar,Poda:2017}. 

Additionally, CUPID will have to deal with the background induced by $\beta$ and $\gamma$'s. Since the intensity of such events drops above the 2615\,keV $\gamma$ peak of $^{208}$Tl, which is generally assumed as the end-point of the environmental $\gamma$ radioactivity, CUPID will search for \DBD\ using an isotope with a higher $Q_{\beta\beta}$ value: $^{100}$Mo, $Q_{\beta\beta}$ = (3034.40$\pm$0.17) keV~\cite{RAHAMAN2008111}.

The combination of scintillating bolometers and high $Q_{\beta\beta}$ value emitters was developed by the LUCIFER~\cite{Beeman:2012jd,Beeman:2012gg,Cardani:2013mja,Beeman:2013sba,Beeman:2013vda,Cardani_2013,Artusa:2016maw,Azzolini_2018} and LUMINEU~\cite{BEEMAN2012318,Barabash:2014una,Armengaud:2015hda,Bekker:2014tfa,Armengaud_2017,Grigorieva2017,Poda2017} projects, as well as by the AMoRE Collaboration~\cite{AMORE2019}.
The outcomes of LUCIFER and LUMINEU were two medium-scale demonstrators, \CupidZ~\cite{PhysRevD.100.092002,PhysRevLett.123.262501,PhysRevLett.123.032501,Azzolini_2019,Azzolini_excited_states,Azzolini_Zn_decay} and CUPID-Mo~\cite{Armengaud:2020:cupidmo,Poda2017,Armengaud2020,Schmidt_2020} respectively. 
Thanks to the high collected statistics, \CupidZ\ proved that the technique of scintillating bolometers allows to suppress the $\alpha$ background by about 3 orders of magnitude, matching the CUPID requirements.
The complementary effort of CUPID-Mo, allowed to assess the performance of \LMO\ crystals in terms of energy resolution, particle identification capability, radio-purity and reproducibility. For these reasons, \LMO\ scintillating bolometer was chosen to be the baseline detector for the CUPID experiment. 

CUPID-Mo used cylindrical \LMO\ crystals coupled to light detectors and surrounded by a reflecting foil. Cubic crystals would largely simplify the construction of a tightly packed array, 
maximizing the emitter mass and enhancing the background suppression via the rejection of coincidences, i.e., events that release energy in more than one crystal.

In this work, we characterized for the first time cubic \LMO\ crystals, in order to prove that they comply with the CUPID goals: an energy resolution of 5\,keV FWHM and complete rejection of the $\alpha$ background in the region of interest.

We have also assessed the impact of a reflecting foil on light collection. Being a potentially contaminated material, the reflecting foil is not a desirable component of the detector. Furthermore, it limits the study of coincidences among crystals, absorbing the $\alpha$ and $\beta$ particles emitted on their surfaces. On the other hand, the light collection that can be achieved without a reflector  was never measured. For this reason, we operated crystals both with and without a reflecting foil.


\section{The 8-crystal prototype}
\label{hardware}
\begin{figure}[bht]
\begin{centering}
\includegraphics[width=\columnwidth]{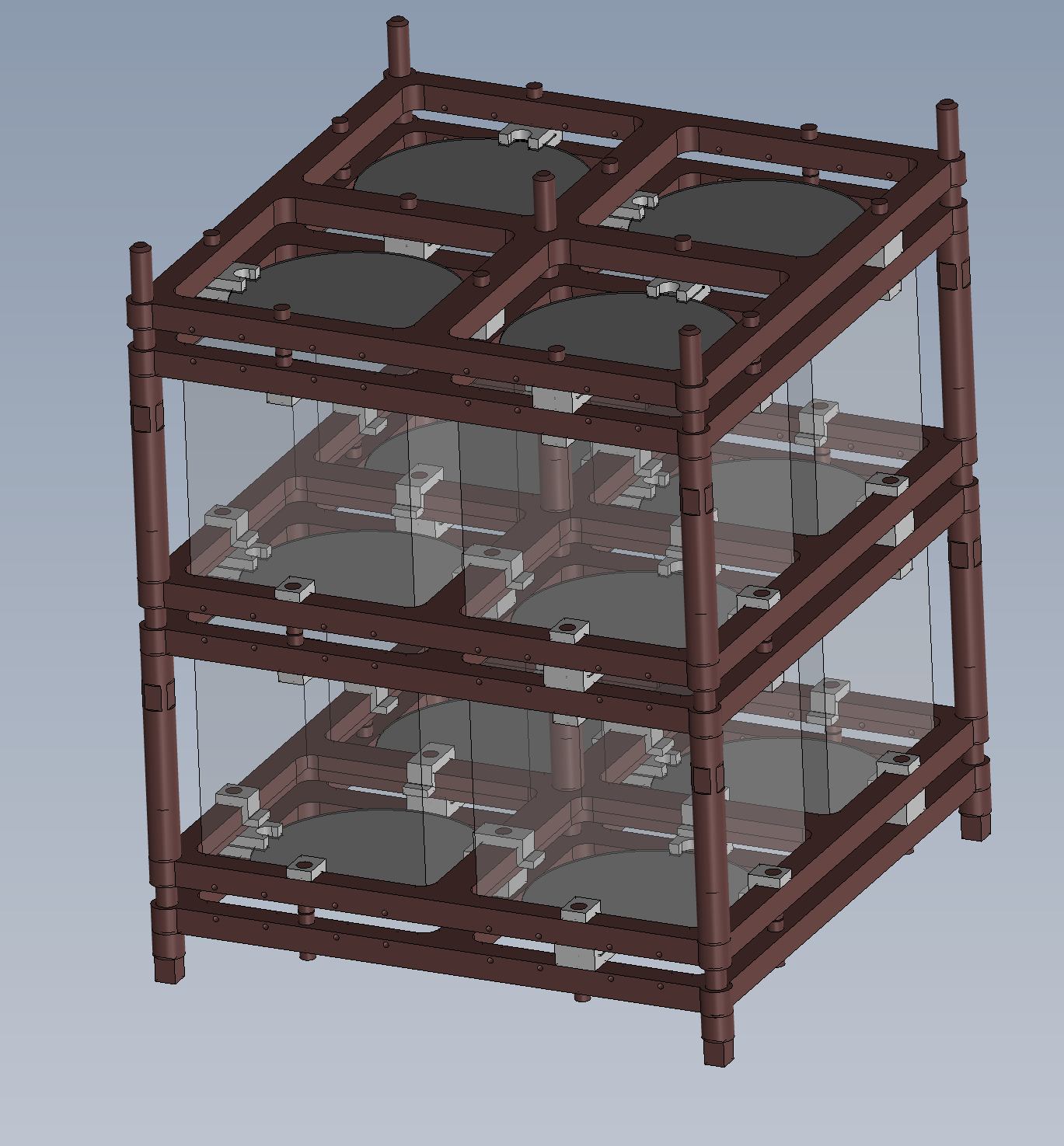}
\caption{Rendering of the 8-crystal array. The array consists of 8 \LMO\ crystals (45$\times$45$\times$45\,mm, corresponding to a mass of $\sim$280\,g) arranged in two floors. Crystals belonging to the bottom floor were surrounded by a reflecting foil (not shown in the rendering). The light emitted by \LMO\ crystals was detected by Ge light detectors ($\oslash$44$\times$0.175\,mm) arranged in three floors.}
\label{figure:setup1}
\end{centering}
\end{figure}

A prototype was designed to fulfill the following requirements:
\begin{itemize}
    \item compact architecture with high efficiency of space usage, as the available space is limited by the experimental volume of the cryostat;
    \item simple and modular assembly, minimizing the number of structural parts;
    \item minimization of the support structure volume and weight;
    \item low radioactivity of all the elements.
\end{itemize}
We designed a prototype consisting of eight \LMO\ crystals disposed in two floors and interleaved by light detectors (Figure~\ref{figure:setup1}). The crystals on the bottom floor were surrounded by a Vikuiti{\texttrademark} from 3M reflecting foil, while those on the top floor were not surrounded by a reflector. 

The mechanical structure included PTFE supports and two types of copper elements: frames and columns. Both materials are available with high radiopurity. Such a structure offers simplicity in assembly and could represent the starting point for the design of the final CUPID towers.

The \LMO\ crystals are (97.7$\pm$0.3)$\%$ enriched in $^{100}$Mo and have a cubic shape with a 45\,mm side and mass of $\sim$280\,g. The growth of the eight Li$_2$MoO$_4$ crystals containing molybdenum enriched in the isotope $^{100}$Mo was performed within the CROSS project~\cite{Bandac2020}, following the protocol set
up by LUMINEU ~\cite{Armengaud_2017,Grigorieva2017}. Due to restrictions imposed by the dimensions of the crucible used for the crystals growth, the edges of the Li$_{2}${}$^{100}$MoO$_4$ samples  were rounded.
The heat signal produced by each \LMO\ crystal was measured using a Neutron Transmutation Doped germanium thermistor (NTD-Ge~\cite{Hallerf}), which produces a typical voltage signal of tens-hundreds of $\mu$V per 1\,MeV of deposited energy. 
Each detector was also equipped with a silicon heater. Periodic reference pulses can be injected via the heater and this is generally used to correct small temperature variations and drifts during the data taking~\cite{Arnaboldi:2003yp,Andreotti2012}. In this characterization run, the injected pulses were instead utilized for pile-up studies on the \LMO\ crystals~\cite{cupid:pileup}.
The sensors and heaters were attached to the crystals using a two-component epoxy resin
(Araldite Rapid{\textregistered}) which is a well studied glue for bolometric detectors. The electrical connections were done through copper pins crimping, as e.g. in CUPID-0~\cite{Azzolini_2018}.

\begin{figure}[thb]
\begin{centering}
\includegraphics[width=\columnwidth]{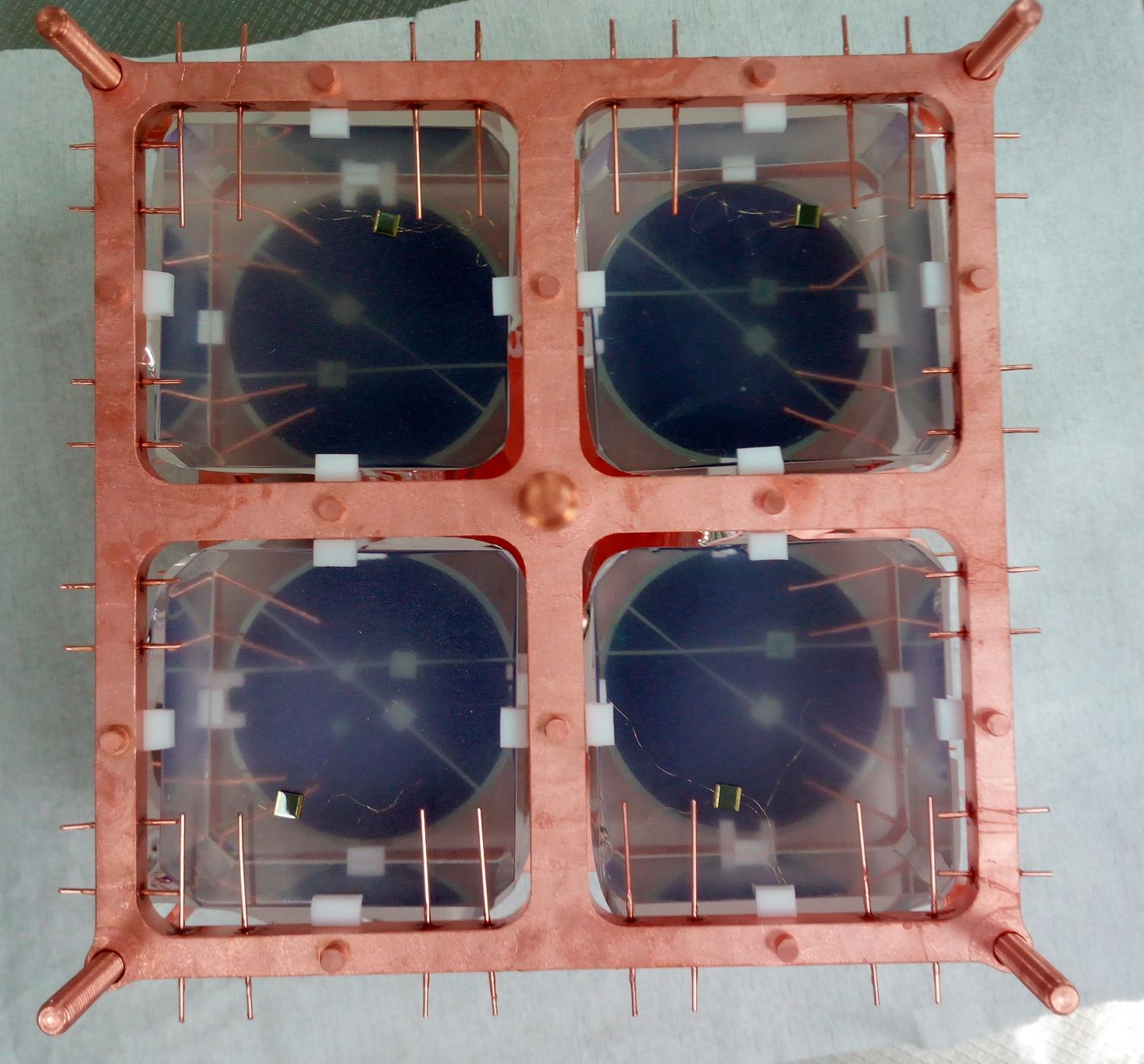}
\caption{Photo of the array during assembly (top view). An NTD-Ge thermistor is glued on top of each crystal. As \LMO\ crystals are almost transparent, we see also the bottom layer of LDs with a SiO coating.}
\label{figure:setup}
\end{centering}
\end{figure}

Concerning the light detectors (LDs), for this work we used the same type of devices already leveraged by CUPID-Mo~\cite{Armengaud2020} and \CupidZ~\cite{Beeman_2013}. 
When an electron of few MeV interacts in a \LMO\ crystal, the expected light signal is of the order of few keV \cite{Armengaud_2017,Poda2017,Armengaud2020}.
``Standard" technologies for light detection (PMT, photodiodes, etc) are not convenient for the applications at cryogenic temperatures. To convert few keV into a readable voltage signal at $\sim$10 mK, we used cryogenic calorimeters made of $\oslash$44$\times$0.175\,mm germanium disks as light detectors. Also the Ge crystals were equipped with an NTD-Ge thermistor and a heater.
To increase the light collection, an antireflecting $\sim$70\,nm SiO layer~\cite{Mancuso2014} was deposited on both sides of LD as e.g. in CUPID-Mo~\cite{Armengaud2020}.
The surface of the LDs foreseen for the thermal sensors and the heater was left uncoated (Figure~\ref{figure:setup}).

In order to calibrate the LDs in the energy scale of  scintillation light signals, we deposited an X-ray source ($^{55}$Fe, emitting  X-rays at 5.9\,keV and 6.4\,keV) on supports facing the surface of the germanium disks.

The prototype was operated in a milli-Kelvin facility located in the Hall C of the deep underground Laboratori Nazionali del Gran Sasso of INFN, Italy.

\section{Data Analysis}
\label{sec:analysis}

The voltage signals were amplified and filtered with a 120 dB/decade, six-pole anti-aliasing active Bessel filter~\cite{Arnaboldi_2018,Carniti2016,Arnaboldi_2008,Arnaboldi2015,Arnaboldi2010,Arnaboldi:2004jj,AProgFE}. We used a custom DAQ software package to save on disk the data stream acquired through a 18 bit analog-to-digital board with a sampling frequency of 2~kHz~\cite{DiDomizio:2018ldc}. 

A derivative trigger was applied to the data to identify thermal pulses; the trigger parameters were tuned for each detector according to its noise and the shape of the thermal pulse response. Moreover, a random trigger was set every 60\,s, to sample detector baselines without any signal. For each trigger, we constructed a 5-s-long time window for both LDs and \LMO\ crystals (from now on, LMO). 

The triggered data were then processed offline via a dedicated analysis chain, which was adapted from a C++ based analysis framework developed for CUORE~\cite{Alduino_analysis_2016}, CUPID-0~\cite{Azzolini:analysis:2018} and their predecessors~\cite{Andreotti_2011}.

A matched filter algorithm (Optimum Filter) ~\cite{Gatti:1986cw,Radeka:1966} was applied to the thermal pulses
to evaluate the signal amplitude by suppressing the most intense noise frequencies.
For each event, we reconstructed other basic parameters, such as the baseline value, which was treated as a proxy for the detector temperature, the baseline RMS, the pulse rise and decay times, and other shape parameters.

We acquired calibration runs with $^{232}$Th sources outside the cryostat, to characterize the LMO  response as a function of the deposited energy. 
The thermal instabilities of the detector could affect its intrinsic gain, resulting in a degradation of energy resolution.  
We used constant energy events (2615\,keV $\gamma$ quanta) to trace the evolution of the pulse amplitude as a function of the crystal temperature, and correct for such dependency (\textit{stabilization} procedure). 

The corrected amplitudes were converted into energy using the most intense gamma peaks produced by the $^{232}$Th sources and building calibration functions accordingly.

We followed a different approach for the LD pulse reconstruction, accounting for their worse signal-to-noise ratio.
The signal template for LDs was built by averaging pulses produced by the $^{55}$Fe X-ray sources. Using this signal template, we applied the Optimum Filter algorithm also to LDs. 
We further improved the estimation of the light amplitude by using the fixed time-delay between light and heat signals due to the jitter of the electronics chain~\cite{Piperno:2011fp}. We first selected high energy events to derive the jitter for each detector. We then evaluated the amplitude of the filtered light pulse at a fixed time delay with respect to the heat pulse.
We underline that such procedure does not change significantly the evaluation of the light signal in the region of interest, but it allows to remove a small non-linearity due to the optimum filter at very low energy.

The energy resolution of LDs is $\sim$1$\%$, thus larger compared to the typical gain variations. 
Thus, we did not have to repeat the stabilization procedure used for the LMO detectors. We energy-calibrated the amplitudes of the LDs using a linear function with zero intercept. The calibration coefficient was derived fitting the 5.9\,keV peak of the $^{55}$Fe source. 

This procedure allowed us to compute the ``light yield" ($LY$) of our detectors, i.e., the amount of scintillation light (in keV) detected in the LDs for a given energy deposition (in MeV) in the LMOs.

\section{Results}
\label{sec:results}

Among the 8 LMO crystals of the presented setup, we noticed that 2 of them (LMO-2, LMO-8) were not functioning when we reached base temperature, due to broken electric contacts. We will present the analysis of the data acquired for the other 6 LMO detectors and related LDs.

The detector response depends on the operating temperature: a lower temperature decreases the thermal capacitance and enhances the NTD-Ge response, generally leading to a better signal-to-noise ratio.

For this work, we operated the detectors at 18\,mK, with working NTD-Ge resistances of 10--50\,M$\Omega$ for the LMO detectors. Due to technical problems related to the test cryogenic facility, it was not possible to operate the detector at $\sim$10\,mK (working resistance of hundreds of M$\Omega$), where we expect the best performance.
We measured a signal amplitude ranging from 30 to 120\,$\mu$V/MeV (mean: $\sim$50 $\mu$V/MeV), and a baseline energy resolution of  1.0--1.8\,keV RMS (mean: $\sim$1.3 keV).

The characteristic rise and decay times (defined as the time difference between the 90\% and the 10\% of the leading edge, and the time difference between the 30\% and 90\% of the trailing edge, respectively) were $\sim$18\,ms and $\sim$120\,ms.
The response of these detectors is consistent with the one of  cylindrical LMO detectors operated in similar conditions~\cite{Armengaud_2017,Armengaud2020}. 

On average, the LDs intrinsic signal amplitude resulted $\sim$5\,mV/MeV, the baseline resolution $\sim$40\,eV RMS and the characteristic times  $\sim$3\,ms (rise) and $\sim$16\,ms (decay). The measured parameters were in full agreement with the typical performance of germanium LDs with NTD-Ge readout~\cite{Artusa:2016maw,Armengaud2020}.

For each LMO channel, we evaluated the energy resolution at the several $\gamma$ peaks used for calibration and extrapolated to the $Q_{\beta\beta}$ of $^{100}$Mo.
Considering all the 6 LMO detectors, the average FWHM at 2615 keV $\gamma$ peak is (7.5$\pm$0.4) keV, while the FWHM at $Q_{\beta\beta}$ is (8.2$\pm$0.5) keV.
However, three LMO detectors (LMO-3, LMO-4, LMO-7) were instrumented with silicon heaters used for pile-up studies~\cite{cupid:pileup}; we observed that the heater system on those LMO crystals was inducing extraneous noise and instabilities in the detectors, affecting the quality of the calibration data. The system to artificially inject pile-up events will not be present in CUPID. Thus, for a more realistic estimation of the CUPID detector performance, we discarded these 3 LMO crystals from the final resolution results. 

\begin{figure}[!thb]
\begin{centering}
\includegraphics[width=\columnwidth]{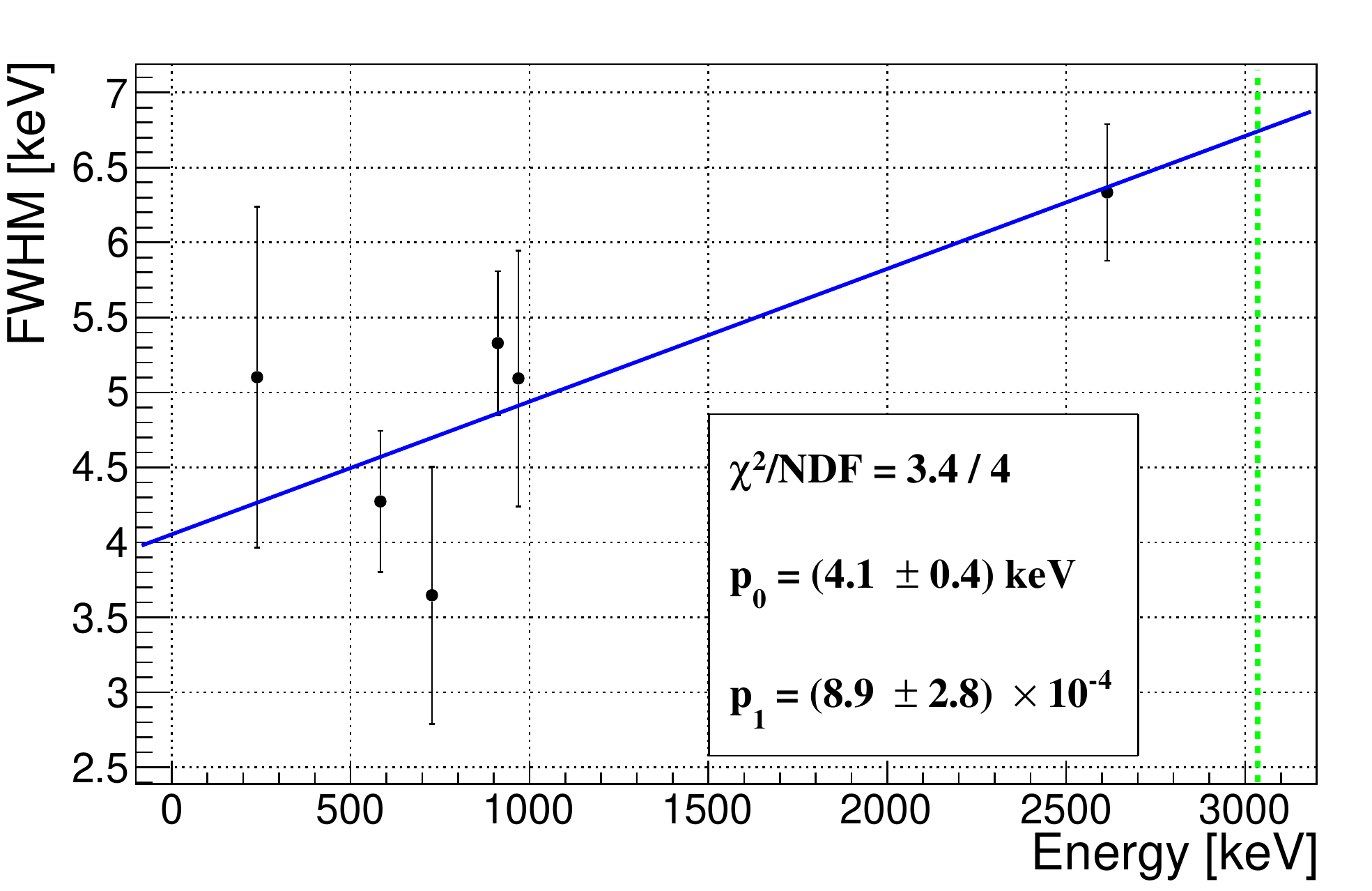}
\caption{FWHM energy resolution as a function of the energy. Data were fitted with a linear function: $FWHM = p_0+p_1\times E$ (blue line). Green dotted line (color online): Q$_{\beta\beta}$ of $^{100}$Mo.}
\label{figure:resoPlot}
\end{centering}
\end{figure}

In Figure~\ref{figure:resoPlot}, we report the FWHM energy resolution as a function of the energy, evaluated from the cumulative calibration spectrum of the other LMO detectors (LMO-1, LMO-5, LMO-6). The extrapolated FWHM at $Q_{\beta\beta}$ is (6.7$\pm$0.6) keV.
This value is already very close to the CUPID goal of 5\,keV FWHM. Furthermore, it has to be interpreted as a conservative value, as the detectors were operated at a higher than expected temperature (limiting the signal-to-noise ratio) and in unstable noise conditions because of major safety upgrades to the cryogenic facility.
The operation at colder temperatures and in stable conditions will allow to further improve this result (as demonstrated with cylindrical LMOs~\cite{Armengaud_2017}).

As mentioned before, one of the main goals of this measurement was to compare the average $LY$ for LMO detectors with/without the reflecting foil and study the discrimination power for $\alpha$ and $\beta/\gamma$ separation in the combined light-heat scenario. 

\begin{figure}[!thb]
\begin{centering}
\includegraphics[width=\columnwidth]{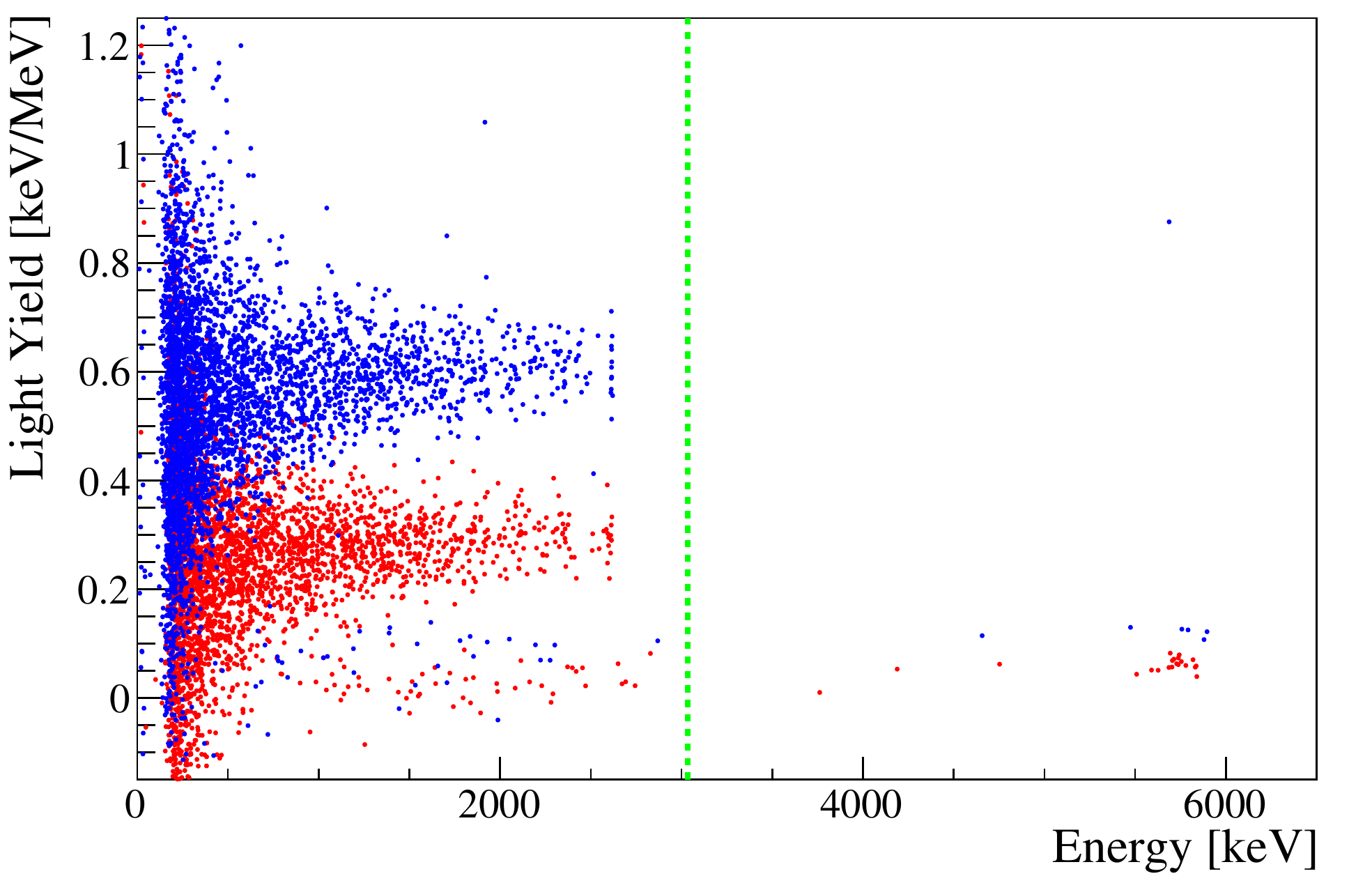}
\caption{Light yield measured in a LD as a function of the energy deposited in the LMO surrounded by the reflecting foil (Blue) and LMO without reflector (Red). Data were collected with a $^{232}$Th $\gamma$ source and a smeared $\alpha$ source.}
\label{figure:LYPlot}
\end{centering}
\end{figure}

Figure~\ref{figure:LYPlot} reports the $LY$ measured by a single light detector as a function of the heat for an LMO surrounded by reflecting foil, and for an LMO without reflector.

In both scenarios, we can clearly identify two populations of events. With an energy reaching 2615\,keV, $\beta/\gamma$ events populate the band with higher $LY$. At very low energy, the light-amplitude can be slightly mis-reconstructed, due to noise superimposed to small scintillation light signals, causing a spread in the $LY$ distribution.

The band with lower $LY$ is populated by $\alpha$ particles.
At high energy we recognize a cluster of events due to an internal crystal contamination in $^{210}$Po (ascribed to a $^{210}$Pb contamination~\cite{Armengaud:2015hda}). This emitter should produce a peak at $\sim$5.4 MeV but, since the detector was energy-calibrated using gamma's, the $\alpha$ peak is observed at slightly higher electron-equivalent energy ($+$7$\%$, in agreement with previous studies with lithium molybdate bolometers~\cite{Cardani_2013,Bekker:2014tfa,Armengaud_2017,Armengaud2020,clymene:2018}).

The other events at lower energies can be ascribed to $\alpha$ particles produced by a $^{238}$U/$^{234}$U source. The source was covered with a thin mylar foil to smear the energy of the $\alpha$ particles and characterize the rejection of the $\alpha$ background also at low energy.

To evaluate the $LY$, we selected electrons and $\alpha$ particles with energy exceeding 1\,MeV. We summed the light collected by LD located on the top and bottom of each LMO crystal, and derived the mean of the distribution of these events. 
We obtained a summed average $LY_{\beta/\gamma}$ = (1.10$\pm$0.05)\,keV/MeV for the LMO surrounded by the reflecting foil, and $LY_{\beta/\gamma}$ = (0.50$\pm$0.05)\,keV/MeV for the LMO without reflector. The $LY_{\alpha}$ for $\alpha$ particles resulted $LY_{\alpha}$ = (0.19$\pm$0.01)\,keV/MeV for the LMO surrounded by the reflecting foil, and $LY_{\alpha}$ = (0.085$\pm$0.004)\, keV/MeV for the LMO without reflector.

The measured $LY$ is lower than that of e.g. cylindrical LMOs of CUPID-Mo (1.35\,keV/MeV for 2 LDs~\cite{Armengaud2020}) because of a non optimal light collection efficiency (more details in the following).
Also, $LY$ differences among LMO crystals were very small (lower than 20$\%$) and mainly due to systematic uncertainties in the energy calibration of the light detectors. 
We finally highlight that, due to the geometry of the array, the top and bottom LDs had the same light collection efficiency. As shown in Figure~\ref{figure:LYPlot}, in which we reported the top LD only, each LD was measuring half of the total collected light.\\

In order to quantify the discrimination capabilities between the $\alpha$ and the $\beta/\gamma$ populations provided by the scintillation signal, we measure the difference between the average $LY$ for signals produced by the two kinds of particles ($LY_{\beta/\gamma}$, $ LY_{\alpha}$). The difference is then compared accounting for the width of the two distributions. 

For this purpose, we defined a Discrimination Power (DP)~\cite{Arnaboldi2011797}:
\begin{equation}
		DP \equiv \frac{\left| LY_{\beta/\gamma} - LY_{\alpha} \right|}{\sqrt{\sigma_{\beta/\gamma}^2 + \sigma_{\alpha}^2}}.
		\label{eq:discr_power}
\end{equation}
We underline that such parameter depends on the energy, as the resolution of LDs (and thus the DP) improves significantly at higher energy (Figure~\ref{figure:LYPlot}). Due to the limited statistics we had to enlarge the region to compute the DP down to 1\,MeV, a region well below the $Q_{\beta\beta}$. As a consequence, the DP values obtained in this work have to be considered as conservative results.

We evaluated the DP for each single LD (top and bottom) and for the sum of the $LY$ of the top-bottom LDs. The results are summarized in Table~\ref{Table:DP-features}. 

\begin{table}[htbp]
\centering
\caption{Discrimination Power (DP) for LD top, LD bottom and the sum of the two light detectors. LMO-1,3,4 are surrounded by the reflecting foil, while LMO-5,6,7 are without a reflecting foil.}
\begin{tabular}{lccc}
\hline
                       &   DP (top)                 &  DP (bottom)  &   DP (sum)        \\
\hline
\hline
LMO-1           & 7.3                            & 8.6             	 & 10.8          \\
\hline
LMO-3        & 6.9                              & 7.1             	    & 10.2       \\
\hline
LMO-4          & 6.9                               & 7.1                & 8.7        \\
\hline
\hline
LMO-5         & 3.4                             & 4.3            	  & 4.6           \\
\hline
LMO-6         & 2.3                               & 3.9               	  & 4.4       \\
\hline
LMO-7         & 4.7                               & 3.3                & 5.5        \\
\hline
\end{tabular}
\label{Table:DP-features}
\end{table}

In case of LMO crystals surrounded by the reflective foil, the average DP for all single LDs is $\sim$7.3, ensuring a complete rejection of $\alpha$ events. In this case, using a second LD would not drastically improve the suppression of the $\alpha$ background, which is already beyond the CUPID target.

The average DP for single LDs facing bare crystals is $\sim$3.6, leading to $\sim$0.05\% of $\alpha$ events which cannot be distinguished from the $\beta/\gamma$ ones.
Even if this rejection capability would comply with the CUPID requirements, we observe that there are some cases in which the DP shows slightly lower values because of higher noise of the LD (for example, for LMO-6 the DP using the top light detector is only 2.3). It is worth observing that summing the light collected by the top/bottom LDs allows to reach a DP much larger than 4, and thus a negligible $\alpha$ background.

We can conclude that with the reflecting foil even a single LD is sufficient for CUPID $\alpha$ discrimination goal, while without a reflecting foil it appears that having two LDs is more effective for tagging the $\alpha$ events even in noisy light detectors.
Summing the $LY$ of the two LDs facing an LMO detector, indeed, can help improve the discrimination capabilities in case of the bare crystals, where the detected scintillation light is poor and so is the single detector $LY$.

The measurement of the $LY$ was used to validate a Monte Carlo simulation of the scintillation light based on the Litrani software~\cite{GENTIT200235}.
We reconstructed the geometry of the detector, using the optical properties of lithium molibdate crystals~\cite{velazquez:hal-01554702}, its scintillation spectrum~\cite{SPASSKY2015195} and scintillation decay time at cryogenic temperatures~\cite{Casali:2019baq}. The optical properties of the coated germanium disk and the reflecting foil were taken from~\cite{Casali_2017}.
The only missing parameter for the simulation was the absolute number of photons emitted by LMO at 18\,mK. To our knowledge, this value is not reported in literature and would require dedicated measurements.
The lack of the number of emitted photons prevented a direct comparison between the simulated and the measured light absorbed by the germanium LDs.
Nevertheless, we could use the simulation to do a comparative study and predict the effect of the reflecting foil. 

\begin{figure}[thb]
\begin{centering}
\includegraphics[width=\columnwidth]{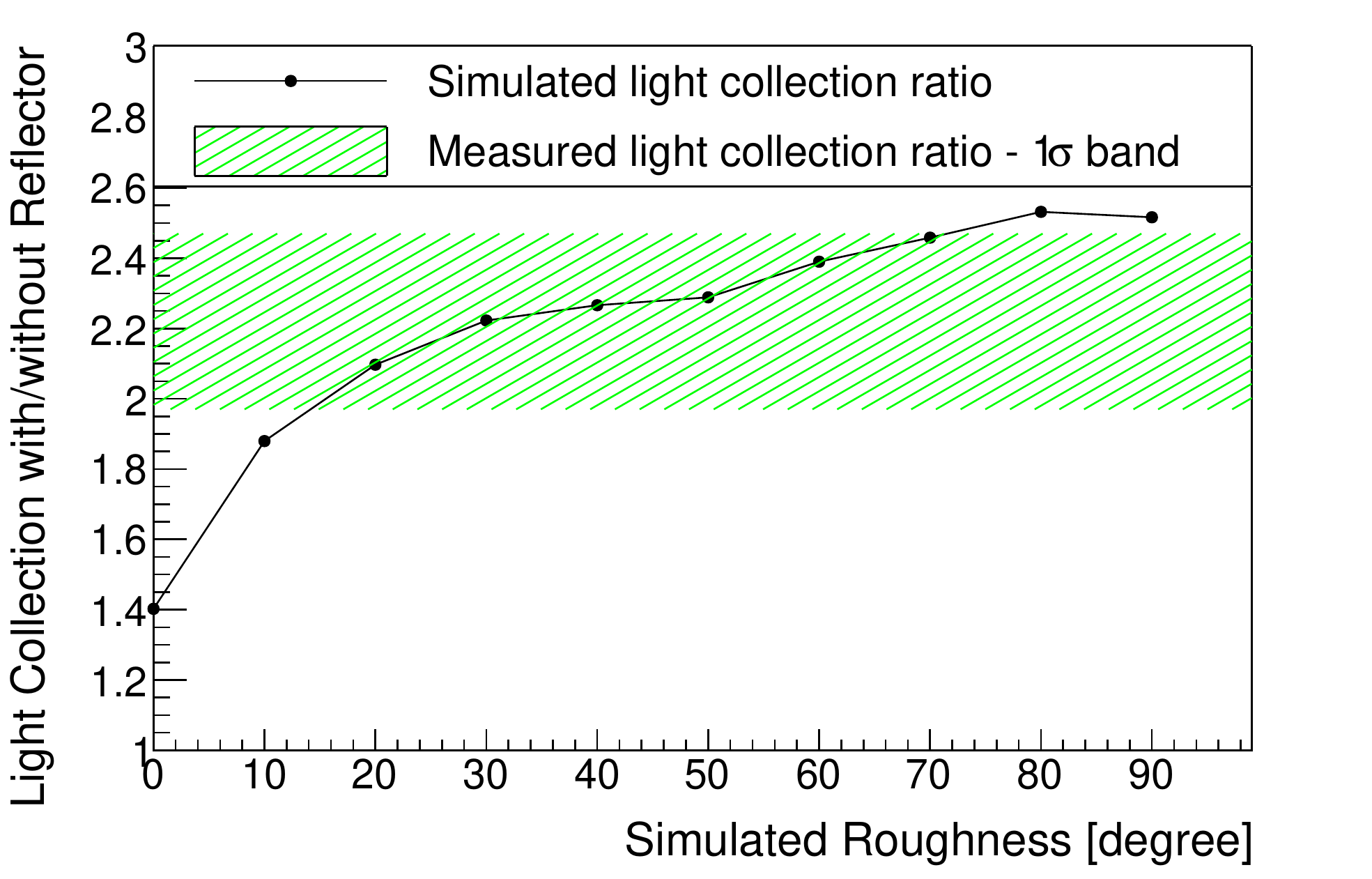}
\caption{Ratio of the light collected with a reflecting foil to the light collected without a reflecting foil, as a function of the surface roughness. The LMO crystals were non optically polished, so the results of interest for this work are those in the central region of the X-axis.}
\label{figure:simulation}
\end{centering}
\end{figure}

In Figure~\ref{figure:simulation}  we report the predicted ratio of the light collected with a reflecting foil to the light collected without a reflecting foil. 

The ratio is reported as a function of the surface roughness, which played an important role with other crystals~\cite{Casali_2017}. In this study, on the contrary, we observed that the emission of scintillation light in LMO crystals does not depend significantly on the surface roughness. The reason for this behaviour resides in the small refractive index of LMO, leading to a narrow critical angle and, as a consequence, to a high escape probability of the emitted photons.

The roughness angle is a simulation parameter that cannot be easily related to the crystal surface. Low  roughness angles describe the scenario in which the crystals are optically polished, while high roughness angles refer to rough crystal surfaces. The crystals used in this work were polished following the LUMINEU protocol, but not to optical grade. As so, we expect the intermediate roughness angles to better describe the experimental conditions. 
We also observe that, especially in this parameter region, the ratio with/without reflector is in good agreement with the measured value of 2.20$\pm$0.24. 

This study shows that the simulation can be an effective tool to model light production, propagation and absorption in scintillating bolometers.

After validating the simulation framework, we used it to predict the improvement in light collection that could be obtained by using (i) a squared light detector that fully covers the LMO side and (ii) a light detector very close to the scintillator. 
In the current prototype, the distance between LMO and light detector is 6.5\,mm. With a different mechanical structure we could narrow it down to 0.5\,mm. 
The simulation suggests that these simple geometrical modifications will enhance the collected light by 60\%.

Preliminary works also proved that putting the light detector in contact with the main crystal allows to increase the light collection without affecting the bolometric performance of the device~\cite{Barucci_2019}.
This experimental scenario cannot be described by a simulation, which would assume an ideal contact between the two surfaces. In reality, both surfaces feature micro-defects preventing an optical contact, which cannot be easily implemented in a simulation.
For this reason, we believe it is important to repeat the studies performed in Ref.~\cite{Barucci_2019} with \LMO\ crystals, and determine if the direct coupling can further enhance light collection.

As proved in this work, increasing the light collection would not be strictly necessary for CUPID. Nevertheless, it would bring many advantages such as:
\begin{itemize}
    \item improving the rejection of background induced by pile-up. Light pulses are faster compared to heat pulses and thus could help disentangling two partly overlapped pulses. Improving the signal-to-noise ratio of such pulses would make the pile-up rejection more effective~\cite{Chernyak:2012,Chernyak:2014,Chernyak:2017};
    \item tolerating the (potential) malfunctioning of one of the two light detectors coupled to the LMO crystals;
    \item coping with a possible higher noise of some LD.
\end{itemize}  
The proposed improvements of the geometry will be studied in future measurements to develop a risk mitigation strategy.

\section{Conclusions}
We tested an array of eight enriched cubic \LMO\ crystals (280\,g each) at the Gran Sasso underground laboratory, in the framework of the CUPID project. Despite the sub-optimal measurement conditions, we obtained an energy resolution of (6.7$\pm$0.6) keV FWHM at $Q_{\beta\beta}$ of $^{100}$Mo (3034 keV), almost in compliance with the CUPID goal of 5\,keV. We foresee a further improvement by operating the crystals at temperatures lower than 18\,mK (reached in the present study) and in more stable cryogenic conditions.

We characterized the particle identification capabilities in different experimental conditions (with and without reflecting foil and using one or two light detectors) and demonstrated  that the baseline design of CUPID already guarantees the necessary rejection of the $\alpha$ background. 

We developed a Monte Carlo simulation of the scintillation light and validated it against the collected data. We identified the limits of the mechanical assembly used in this test and proposed some upgrades for the structure of the CUPID towers. According to the simulation, such improvements will allow to enhance the light collection by 60\%. 
Even if increasing the light collection is not crucial for CUPID, it will bring many advantages both in terms of background suppression and risk mitigation.

\begin{acknowledgements}
The CUPID Collaboration thanks the directors and staff of the
Laboratori Nazionali del Gran Sasso and the technical staff of our
laboratories. This work was supported by the Istituto Nazionale di
Fisica Nucleare (INFN); by the European Research Council (ERC) under
the European Union Horizon 2020 program (H2020/2014-2020) with the ERC
Advanced Grant No.\ 742345 (ERC-2016-ADG, project CROSS) and the Marie
Sklodowska-Curie Grant Agreement No.\ 754496; by the Italian Ministry
of University and Research (MIUR) through the grant Progetti di
ricerca di Rilevante Interesse Nazionale (PRIN 2017, grant
No.\ 2017FJZMCJ); by the US National Science Foundation under Grant
Nos.\ NSF-PHY-1401832, NSF-PHY-1614611, and NSF-PHY-1913374. This material is also based
upon work supported by the US Department of Energy (DOE) Office of
Science under Contract Nos. DE-AC02-05CH11231 and DE-AC02-06CH11357;
and by the DOE Office of Science, Office of Nuclear Physics under
Contract Nos.\ DE-FG02-08ER41551, DE-SC0011091, DE-SC0012654,
DE-SC0019316, DE-SC0019368, and DE-SC0020423. This work was also
supported by the Russian Science Foundation under grant
No.\ 18-12-00003 and the National Research Foundation of Ukraine under
Grant No.\ 2020.02/0011. This research used resources of the National
Energy Research Scientific Computing Center (NERSC). This work makes
use of both the DIANA data analysis and APOLLO data acquisition
software packages, which were developed by the CUORICINO, CUORE,
LUCIFER and CUPID-0 Collaborations.
\end{acknowledgements}


\end{document}